\documentclass[letterpaper, 10 pt, journal, twoside]{ieeetran}
\IEEEoverridecommandlockouts
\usepackage{cite}
\usepackage[dvipsnames]{xcolor}
\usepackage{amsmath,amssymb,amsfonts,mathtools}
\usepackage{algorithmic}
\usepackage{graphicx}
\usepackage[export]{adjustbox}
\usepackage{tikz}
\usepackage{textcomp}
\usepackage[font=small]{caption}
\def\BibTeX{{\rm B\kern-.05em{\sc i\kern-.025em b}\kern-.08em
		T\kern-.1667em\lower.7ex\hbox{E}\kern-.125emX}}

\usepackage{amsthm}
\usepackage{mathrsfs}
\usepackage{hyperref}
\usepackage{enumerate}

\newcommand{\textsr}[1]{\textrm{\textnormal{#1}}}

\newcommand{\reals}{\mathbb{R}}
\newcommand{\transpose}{^\textsr{T}}

\DeclareMathOperator*{\argmin}{arg\,min}

\DeclareMathOperator{\floor}{floor}
\DeclareMathOperator{\closure}{cl}

\newcommand{\smallin}[0]{\,{\in}\,}
\newcommand{\smallop}[1]{\,{#1}\,}

\newtheorem{theorem}{Theorem}
\newtheorem{assumption}{Assumption}
\newtheorem{lemma}{Lemma}
\newtheorem{definition}{Definition}

\newtheorem{remark}{Remark}

\newtheorem{rules}{Rule}

\usepackage{verbatim}
\usepackage{url}
\usepackage{hyperref}
\usepackage{placeins}
\usepackage{soul}
\usepackage{tikz}

\definecolor{lcolor}{rgb}{0,0,0.6}
\definecolor{edits}{rgb}{0,0,0}
\hypersetup{
	colorlinks=true,
	linkcolor=lcolor,
	filecolor=magenta,      
	urlcolor=black,
}

 \newcommand{\regularversion}[1]{\iffalse{}#1\fi}
 \newcommand{\extendedversion}[1]{{#1}}

\begin{document}

\title{Robust Safety-Critical Control for Systems with Sporadic Measurements and Dwell Time Constraints}%

\author{Joseph Breeden\thanks{J. Breeden is with the Department of Aerospace Engineering at the University of Michigan, Ann Arbor, MI, United States, \texttt{jbreeden@umich.edu}.}, Luca Zaccarian\thanks{L. Zaccarian is with the Laboratoire d’Analyse et d’Architecture des Systèmes---CNRS, Université de Toulouse, UPS, 31400 Toulouse, France, and with the Dipartimento di Ingegneria Industriale, University of Trento, 38122 Trento, Italy, \texttt{luca.zaccarian@laas.fr}.}, and Dimitra Panagou\thanks{D. Panagou is with the Department of Robotics and the Department of Aerospace Engineering at the University of Michigan, Ann Arbor, MI, United States, \texttt{dpanagou@umich.edu}.}\thanks{This work was supported by the Chateaubriand Program of the Office for Science \&
Technology of the Embassy of France in the U.S., the Fran\c{c}ois Xavier Bagnoud Foundation, the MUR via grant STARLIT CUP E53D23001130006 \#2022ZE9J9J, the ANR via grant OLYMPIA \#ANR-23-CE48-0006, and the U.S. National Science Foundation grant \#1942907.}}%

\maketitle

\thispagestyle{empty}

\begin{abstract}
This paper presents extensions of control barrier function (CBF) theory to systems with disturbances wherein a controller only receives measurements infrequently and operates open-loop between measurements, while still satisfying state constraints. The paper considers both impulsive and continuous actuators, and models the actuators, measurements, disturbances, and timing constraints as a hybrid dynamical system. We then design an open-loop observer that bounds the worst-case uncertainty between measurements. We develop definitions of CBFs for both actuation cases, and corresponding conditions on the control input to guarantee satisfaction of the state constraints. We apply these conditions to simulations of a satellite rendezvous in an elliptical orbit and autonomous orbit stationkeeping.%
\end{abstract}

\begin{IEEEkeywords}
Constrained control, hybrid systems, aerospace%
\end{IEEEkeywords}

\section{Introduction}

This paper considers the design of Control Barrier Function (CBF) \cite{CBFs_Tutorial} based safety filters for systems with infrequent measurements, and either impulsive or continuous{} actuators. In the authors' prior work \cite{CDC23}, the authors developed safety filters 
for systems with impulsive actuators and deterministic \regularversion{evolution}\extendedversion{system evolution}. This paper builds on \cite{CDC23} by 1) introducing a timed measurement model and extending \cite{CDC23} to ensure set invariance for \extendedversion{perturbed }systems with bounded disturbances\extendedversion{{ }(equivalently bounded model uncertainty)}, 2) rewriting the dynamics \cite[Eq.~(1)]{CDC23} and the new measurement model as a general hybrid system, and 3) also applying this timed measurement model to set invariance \extendedversion{problems }with continuously-applied actuators.

This work and the prior work \cite{CDC23} are motivated by the challenge of satellite control design. Most satellite actuators are either A) impulsive, causing a (practically) instantaneous change in the satellite velocity \cite{spacecraft_stability}, or B) low-thrust, causing a small change in the{} satellite velocity that accumulates over time \cite{low_thrust_example}. The former is often modelled by impulse differential equations \cite{hybrid_impulse_inclusion} or hybrid systems \cite{hybrid_cbfs_hybrid_systems}, while the latter is modelled by conventional continuous-time dynamics. In this work, we utilize hybrid dynamics, or rather hybrid inclusions, as \extendedversion{this is }a natural way to model both the system uncertainties and the impulsive actuations. Several works \cite{hybrid_cbfs_hybrid_systems,hybrid_synergistic_cbfs,hybrid_robust_part_ii,adding_obstacles} have considered CBF-like conditions for hybrid \extendedversion{dynamical }systems, but \cite{CDC23} was the first to our knowledge to consider a minimum \emph{dwell time} between actuations. As shown in \cite{CDC23}, the consideration{} of a minimum delay between actuations makes the control synthesis{} problem more similar to that of a sampled-data controller \cite{LCSS,CBFs_Mechanical}, which must satisfy all constraints at least until the next sample time, than to \cite{hybrid_cbfs_hybrid_systems,hybrid_synergistic_cbfs,hybrid_robust_part_ii,adding_obstacles,self_triggered_cbf,event_triggered_journal}. However, the dwell time is often too large to use the bounding approximations in \cite{LCSS,CBFs_Mechanical}, so \cite{CDC23} also uses explicit trajectory predictions, which this work generalizes to tubes of predicted possible trajectories. Using these tubes, the minimum dwell time can be naturally extended to a minimum time between measurements as well.

Such measurement timings{} are important because satellites are generally incapable of measuring their own state without external equipment. \extendedversion{Satellites below the Global Navigation Satellite Systems (GNSS) can make use of those satellites, and some work has been done to extend GNSS availability above the GNSS altitude \cite{sidelobe_gps,pulsar_gps}. However, generally satellites in this regime or outside Earth orbit rely on scheduled ranging measurements from ground stations.}\regularversion{Rather, satellites outside the GNSS regime rely on scheduled ranging measurements communicated from ground stations.}
The capacity of existing ground stations is limited, so satellites may run long periods open-loop before receiving updated measurements{}. \mbox{Measurement-robust} CBFs were studied in \cite{measurement_robsut_cbfs,early_noise_cbfs} and observers with CBFs were considered in \cite{belief_cbfs,observer_cbfs,dev_cbf_observers}. Sampled-data considerations were added in \cite{sampled_data_uncertain_robust_cbfs}, but to our knowledge, no work has considered systems that run open-loop for long durations between measurements using CBFs. By contrast, this work allows for arbitrarily many control samples to occur between consecutive measurements.

This paper is organized as follows. Section~\ref{sec:impulsive_model} presents the impulsive actuation model and the{} observer design with minimum dwell time between actuations and bounded delays between measurements. Section~\ref{sec:safety_impulsive} presents the safety-preserving control design for this model. Section~\ref{sec:continuous} presents related safety conditions for continuously-applied actuators. Section~\ref{sec:simulations} presents simulation case studies. Finally, Section~\ref{sec:conclusions} presents concluding remarks. \regularversion{All theorem proofs are found in the extended version \cite{extended_version}.{}} We note that while this work is motivated by satellite control, the following formulations may also be relevant to other infrequently-measured systems.

\noindent%
\textbf{Notations:}
Let $\mathcal{B}_r(x)$ denote the unit ball of radius $r$ centered at $x$. To avoid excessive notation, the domains and ranges of most functions in this paper are implied. 
All norms are the two-norm. Let $\mathcal{K}$ be the set of extended class-$\mathcal{K}$ functions.

\section{Impulsive Control and Measurement Models} \label{sec:impulsive_model}

\subsection{Impulsive Control Explanation}

\begin{figure}
    \centering
    \resizebox{0.88\columnwidth}{!}{\begin{tikzpicture}
        \node[anchor=south west,inner sep=0] (image) at (0,0) {\includegraphics[width=0.95\columnwidth]{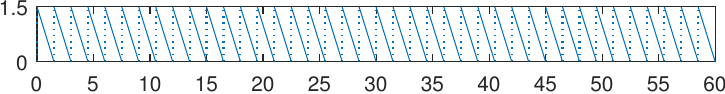}};
        \node[anchor=south west,inner sep=0] (image) at (0.05,1.16) {\includegraphics[width=0.935\columnwidth]{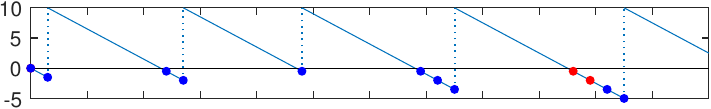}};
        \node[anchor=south west,inner sep=0] (image) at (0.025,2.49) {\includegraphics[width=0.938\columnwidth]{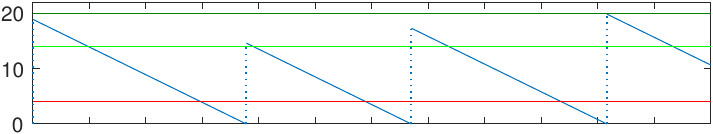}};
        \node [anchor=south west, rotate=90] (note) at (0.1,0.45) {$\sigma_s$}; 
        \node [anchor=south west, rotate=90] (note) at (0.1,1.57) {$\sigma_a$}; 
        \node [anchor=south west, rotate=90] (note) at (0.1,2.92) {$\sigma_m$}; 
        \node [anchor=south west] (note) at (8.27,2.59) {\color{BrickRed}$T_m$}; 
        \node [anchor=south west] (note) at (8.27,3.22) {\color{green}$T_L$};  
        \node [anchor=south west] (note) at (8.27,3.6) {\color{OliveGreen}$T_M$};
        \node [anchor=south west] (note) at (8.27,2.05) {$T_a$};
        \node [anchor=south west] (note) at (8.27,0.73) {$T_s$};
    \end{tikzpicture}}
    \caption{Visualization of the three timers (same units for both axes).}
    \label{fig:timers}
\end{figure}
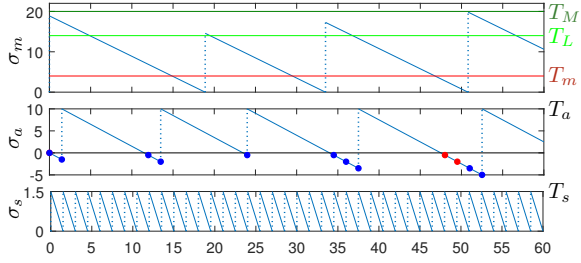

In this section, we present the impulsive actuation and measurement models that we \extendedversion{will }later control\extendedversion{{ }in Section~\ref{sec:safety_impulsive}}. We suppose that the satellite possesses only impulsive actuators, and impose several rules on the timing of these actuation impulses, which \regularversion{are}\extendedversion{will be} incorporated into three timers $\sigma_s,\sigma_a,\sigma_m$\extendedversion{{ }in the model \eqref{eq:model_hybrid}. These rules are as follows and a visual explanation is shown in Fig.~\ref{fig:timers}}\regularversion{, shown in Fig.~\ref{fig:timers}}.%

\begin{rules} \label{rule:sample}
    The controller is sampled with fixed period $T_s\in\reals_{>0}$ and a control impulse can only be applied at the sample times.
\end{rules}

\noindent
That is, the control law computation requires finite time, 
so we assume that{} these computations occur at a fixed rate. This is encoded in the \emph{sampling timer}, $\sigma_s \in [0, T_s]$. Note that $\sigma_s$ is for control sampling, not measurement sampling. An actuation can occur only if $\sigma_s = 0$, as shown in the bottom plot of Fig.~\ref{fig:timers}.

\begin{rules} \label{rule:actuator}
    A control impulse can only be applied at least $T_a \in \reals_{> 0}$ after the last nonzero control impulse.
\end{rules}
\noindent
That is, the actuators have a ``cool-down'' period or dwell-time between uses. This is encoded in the \emph{actuation timer}, $\sigma_a\in(-\infty, T_a]$. An actuation can occur only if $\sigma_a \leq 0$, as indicated by the red and blue dots in the center plot of Fig.~\ref{fig:timers}.

Rules~\ref{rule:sample}-\ref{rule:actuator} also appeared in our prior work \cite{CDC23}. New to this work, we now consider model and measurement uncertainty. In practice, the procedure to measure a satellite's state is to take several range measurements over a time-interval and then to compute a curve fitting. This implies that 1) measurements must be scheduled and coordinated with the ground station, 2) there may be a long delay between measurements during which the satellite runs 
open-loop, and 3) the satellite must not apply any control during the measuring time-interval so as to not corrupt the measurement, as captured in the following rule.

\begin{rules} \label{rule:measure_delay}
    A control impulse cannot be applied in the $T_m \in \reals_{\geq 0}$ time interval before the next measurement.
\end{rules}
\noindent
That is, $T_m$ is the time-interval required to complete the measurement.
To model this, we introduce the \emph{measurement timer}, $\sigma_m\in[0,T_M]$, where $T_M$ is explained in Rule~\ref{rule:measurements} in Section~\ref{sec:measurements} and where $\sigma_m$ is the time until the next measurement. Due to Rule~\ref{rule:measure_delay}, no control input can be applied when $\sigma_m\in[0,T_m)$, as represented by the red dots in Fig.~\ref{fig:timers}.

The timers $\sigma = (\sigma_s,\, \sigma_a,\, \sigma_m)$ evolve in the set $\boldsymbol{\Sigma} \triangleq [0,T_s]\times(-\infty,T_a]\times[0,T_M]$. Any $\sigma$ that satisfies all of Rules~\ref{rule:sample}-\ref{rule:measure_delay} is an \emph{impulse opportunity}, shown by blue dots in Fig.~\ref{fig:timers}, and lies in 
\begin{equation}
    \boldsymbol{\Sigma}_a \triangleq \{ \sigma \in \boldsymbol{\Sigma} \mid \sigma_s = 0 \;\textrm{and}\; \sigma_a \leq 0 \;\textrm{and}\; \sigma_m \geq T_m \}.
\end{equation}%

\subsection{Hybrid Dynamical Model of Impulsive Control}

\begin{figure}
    \centering
    \includegraphics[width=0.76\columnwidth,trim={0in, 2.15in, 0in, 0in},clip]{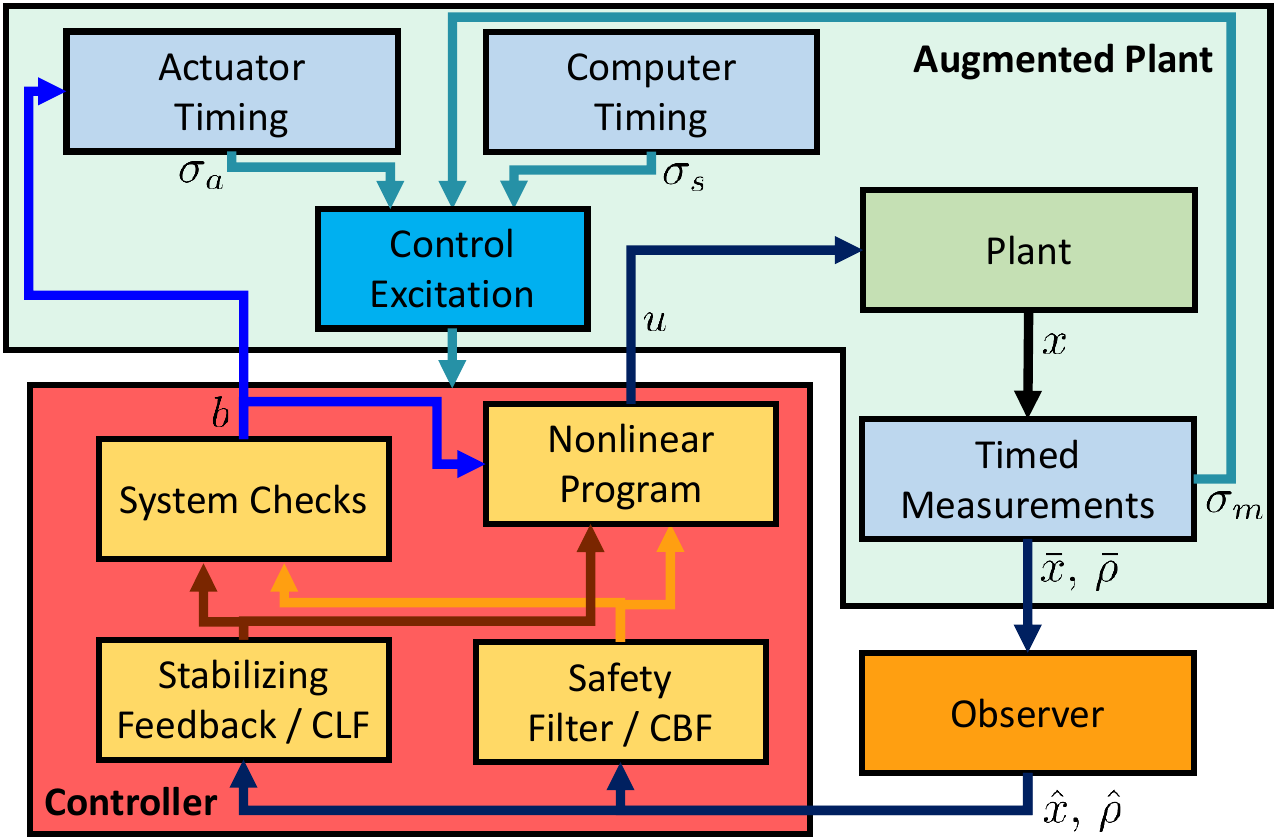}
    \caption{Block diagram of the impulsive system}
    \label{fig:block_diagram}
\end{figure}

We model the impulsive actuator and the above rules with a hybrid dynamical system, visualized fully in Fig.~\ref{fig:block_diagram}. Denote the set of considered times as $\mathcal{T}\subseteq\reals$. Suppose a continuous state $x = (r,v)\in\mathcal{X}(t) \subseteq\reals^{2n}$. Here, $\mathcal{X}$ encodes the set of ``reasonable states'' and is used to define Lipschitz constants later. We assume $x$ is divided into two parts $r,v$ with the second-order structure $\dot{r} \smallop{=} v$, because this will allow for reduced conservatism, as explained later in Remark~\ref{rem:second_order}.

We model Rules~\ref{rule:sample}-\ref{rule:measure_delay} using two control inputs. The first control input, $b\in\{0,1\}$, encodes whether an impulse actually occurs at each impulse opportunity. If $b \smallop{=} 0$, then the controller is allowed to apply an impulse but chooses not to, so the timer $\sigma_a$ is{} not reset. The second control input, $u\in\mathcal{U}\subseteq\reals^m$ models the magnitude of an impulse, if one occurs. Otherwise, we require that $b = 0 \implies u = 0$.

Now, we present the hybrid model. Assume that the system is capable of flowing for all $x\in\mathcal{X}$, so only the timers $\sigma$ and discrete control $b$ are part of the definition of the flow and jump sets. For brevity, denote $z \smallop{=} (b, \sigma)$ and \extendedversion{define }$\mathcal{Z} \smallop{\triangleq} \{0,1\}\times \boldsymbol{\Sigma}$. The first of the three \extendedversion{possible }jump sets is when a measurement occurs:%
\begin{subequations}
\begin{equation}
    \mathcal{D}_m \triangleq \{ z \in \mathcal{Z} \mid \sigma_m = 0\}. 
\end{equation}
Note that, unlike control impulses, measurements always occur when $\sigma_m = 0$, regardless of the other states/timers. The second jump set is when a sample and an actuation impulse occur:
\begin{equation}
    \mathcal{D}_a \triangleq \{ z \in \mathcal{Z} \mid \sigma \in \boldsymbol{\Sigma}_a\;\textrm{and}\; b = 1\}.
\end{equation}
Finally, the third jump set is when a sample occurs but an actuation impulse does not occur:
\begin{equation}
    \hspace{-4pt}\mathcal{D}_c \smallop{\triangleq} \{ z \smallop{\in} \mathcal{Z} \mid \sigma_s \smallop{=} 0 \;\textrm{and}\; (\sigma_a \smallop{\geq} 0 \;\textrm{or}\; \sigma_m \smallop{\leq} T_m \;\textrm{or}\; b \smallop{=} 0)\}.\hspace{-4pt}
\end{equation}
\label{eq:jump_sets}%
\end{subequations}
The flow set is the closure of the complement
\begin{equation}
    \mathcal{C} \triangleq \closure(\mathcal{Z} \setminus (\mathcal{D}_m \cup \mathcal{D}_a \cup \mathcal{D}_c)) .
\end{equation}

We are now ready to write the hybrid dynamics:
\begin{equation}
    \begin{aligned}
    &\begin{cases}
        \dot{r} = v \\ 
        \dot{v} \in F(t,x) \\
        \dot{\sigma}_s = -1 & z \in\mathcal{C} \\
        \dot{\sigma}_a = -1 \\
        \dot{\sigma}_m = -1
    \end{cases}, \hspace{30pt} \begin{cases}
        r^+ = r \\
        v^+ = v \\
        \sigma_s^+ = \sigma_s & z \in \mathcal{D}_m \\ 
        \sigma_a^+ = \sigma_a \\
        \sigma_m^+ = \bar{\sigma}_m(j) \hspace{-10pt}
    \end{cases} \\ &\begin{cases}
        r^+ = r \\
        v^+ \in v + G(t,u) \hspace{-8pt} \\
        \sigma_s^+ = \sigma_s & z\in\mathcal{D}_a \\
        \sigma_a^+ = T_a \\
        \sigma_m^+ = \sigma_m 
    \end{cases}, \hspace{8pt}\begin{cases}
        r^+ = r \\
        v^+ = v \\
        \sigma_s^+ = T_s & z\in\mathcal{D}_c \\
        \sigma_a^+ = \sigma_a \\
        \sigma_m^+ = \sigma_m
    \end{cases}\end{aligned} \label{eq:model_hybrid}
\end{equation}
Here, $F$ and $G$ are our model dynamics, which are set-valued and obey Assumption~\ref{as:dynamics}{} below, and\extendedversion{{ }the discrete input} $\bar{\sigma}_m$ is the time until the next measurement. A solution is parameterized by hybrid time $(t,j)\in\mathcal{T}\times\mathbb{Z}_{\geq 0}$. Assume that $F$ is sufficiently regular that solutions of \eqref{eq:model_hybrid} exist for all time. Note that multiple jumps may occur at a single instant, but Zeno behavior is impossible provided that $\bar{\sigma}_m > 0$ (as will be ensured by Rule~\ref{rule:measurements} below). \extendedversion{The model \eqref{eq:model_hybrid} excludes dependence of $G$ on the state $x$, but this is a straightforward extension.}

\begin{assumption} \label{as:dynamics}
    The dynamics $F,G$ obey the following: 
    \begin{enumerate}
        \item There exists known function $f(t,x)$ and constant $w_c\in\reals_{\geq 0}$ such that $F(t,x) \subseteq \mathcal{B}_{w_c}(f(t,x))$ for all $t,x$.
        \item There exist known functions $g(t,u)$ and $w_g:\reals_{\geq 0}\rightarrow\reals_{\geq 0}$ such that $g(t,0) = 0$, $w_g(0) = 0$, $w_g$ is continuous, and $G(t,u) \subseteq \mathcal{B}_{w_g(\|u\|)}(g(t,u))$. Additionally, let $W_g = \sup_{u\in\mathcal{U}} w_g(\|u\|)$ and assume that $W_g$ is finite.
        \item There exist global Lipschitz constants $\ell_{f,r},\ell_{f,v}\in\reals_{\geq 0}$ such that $\| f(t,(r_1,v_1)) - f(t,(r_2,v_2)) \| \leq\ell_{f,r} \|r_1 - r_2 \| +\ell_{f,v} \| v_1 - v_2 \|$ for all $(r_1,v_1),(r_2,v_2)\in\mathcal{X}(t)$, $t\in\mathcal{T}$.%
    \end{enumerate}%
\end{assumption}%

\noindent
That is, we possess ``best estimate'' functions $f$ and $g$ with bounded uncertainties compared to the real model $F$~and~$G$. Here, $w_c$ might represent modeling error and $w_g$ might represent actuator misalignment. 
We note that both systems in Section~\ref{sec:simulations} satisfy the global Lipschitzness assumption, though this assumption can be relaxed \regularversion{\cite[Remark~3]{extended_version}}\extendedversion{(see Remark~\ref{rem:lipschitz})}.

\subsection{Measurements and Impulsive Open-Loop Observer} \label{sec:measurements}

As explained above, a satellite is often unable to measure its own state, and must rely on measurements communicated from ground stations. We say that a \emph{measurement instant} occurs when $\sigma_m = 0$ and the system jumps according to $\mathcal{D}_m$.

\begin{rules} \label{rule:measurements}
    At each measurement instant, the ground \regularversion{transmits}\extendedversion{station communicates}
\begin{enumerate}[i)]
    \item a new estimate $\bar{x} = (\bar{r},\bar{v})\in\mathcal{X}$ of the satellite's state,
    \item bounds on the estimate error $\bar{\rho} = (\bar{\rho}_r,\bar{\rho}_v) \in [0, \rho_r] \times [0,\rho_v]$ where $\rho_r,\rho_v$ are the maximal errors{} of the station,
    \item the time $\bar{\sigma}_m \in [T_L, T_M]$ to the next measurement instant,
\end{enumerate}
where $\rho_r,\rho_v\in\reals_{\geq 0}$, and $T_L,T_M\in\reals_{>0}$ are known values and where the real state satisfies $r \in \mathcal{B}_{\bar{\rho}_r}(\bar{r})$ and $v \in \mathcal{B}_{\bar{\rho}_v}(\bar{v})$.
\end{rules}

That is, at regular intervals (intervals of length bounded by $T_M$), the observer block in Fig.~\ref{fig:block_diagram} receives updated estimates of the system state and error bounds.
Due to the disturbances, the state knowledge will gradually degrade before the next measurement. Let $\hat{x} = (\hat{r},\hat{v})$ be an estimate of the state and $\hat{\rho} = (\hat{\rho}_r, \hat{\rho}_v)$ be an upper bound on the uncertainty. Assume that the timers are all known perfectly. Consider the following open-loop observer design:
\begin{equation}
    \hspace{-6pt}\begin{aligned}
    &\begin{aligned}\begin{cases}
        \dot{\hat{r}} = \hat{v} \\ 
        \dot{\hat{v}} = f(t,\hat{x}) \\
        \dot{\hat{\rho}}_r = \hat{\rho}_v \\
        \dot{\hat{\rho}}_v = \ell_{f,r} \hat{\rho}_r + \ell_{f,v} \hat{\rho}_v + w_c \hspace{-24pt}
    \end{cases} & z\in\mathcal{C} \end{aligned} \hspace{-9pt}, \hspace{10pt}
    \begin{aligned} \begin{cases}
        \hat{r}^+ = \bar{r}(j) \\
        \hat{v}^+ = \bar{v}(j) \\
        \hat{\rho}_r^+ = \bar{\rho}_r(j) \\
        \hat{\rho}_v^+ = \bar{\rho}_v(j)
    \end{cases} & z\in\mathcal{D}_m \end{aligned} \\
    &\begin{aligned} \begin{cases}
        \hat{r}^+ = \hat{r} \\
        \hat{v}^+ = \hat{v} + g(t,u) \\
        \hat{\rho}_r^+ = \hat{\rho}_r \\
        \hat{\rho}_v^+ = \bar{\rho}_v + w_g(\|u\|)
    \end{cases} & z\in\mathcal{D}_a \end{aligned} \hspace{-9pt}, \hspace{14pt} 
    \begin{aligned} \begin{cases}
        \hat{r}^+ = \hat{r} \\
        \hat{v}^+ = \hat{v} \\
        \hat{\rho}_r^+ = \hat{\rho}_r \\
        \hat{\rho}_v^+ = \hat{\rho}_v
    \end{cases} & z\in\mathcal{D}_c \end{aligned}
    \end{aligned}\hspace{-25pt} \label{eq:observer_hybrid}
\end{equation}

Note that we have constructed \eqref{eq:observer_hybrid} so that the flow dynamics of $\hat{\rho}$ are linear time-invariant

\begin{subequations}\begin{equation}
    \frac{d}{dt}\begin{bmatrix} \hat{\rho}_r \\ \hat{\rho}_v \end{bmatrix} = \underbrace{\begin{bmatrix} 0 & 1 \\ \ell_{f,r} & \ell_{f,v} \end{bmatrix}}_{=A} \begin{bmatrix} \hat{\rho}_r \\ \hat{\rho}_v \end{bmatrix} + \begin{bmatrix} 0 \\ w_c \end{bmatrix} \,.
\end{equation}

Thus, the uncertainties $\hat{\rho}$ have an explicit solution. 
If $\ell_{f,r} > 0$, then this flowing{} solution is (where $I$ is the identity matrix)
\begin{equation}
    \hspace{-4pt}\begin{bmatrix} \hat{\rho}_r(\tau,j) \\ \hat{\rho}_v(\tau,j) \end{bmatrix} = e^{A (\tau-t)} \begin{bmatrix} \hat{\rho}_r(t,j) \\ \hat{\rho}_v(t,j) \end{bmatrix} + A^{-1} (e^{A (\tau-t)} - I) \begin{bmatrix} 0 \\ w_c \end{bmatrix} \hspace{-3pt}. \hspace{-4pt} \label{eq:linear_estimates}
\end{equation}\label{eq:estimates}\end{subequations}%
\vspace{-10pt}

\begin{remark} \label{rem:second_order}
    The authors chose to specify the second order structure of $(r,v)$ in \eqref{eq:model_hybrid} so that the linear system \eqref{eq:estimates} would appear. If one instead treats $\dot{x} = f(x)$ as a general system with a single Lipschitz constant $\ell_f\in\reals_{\geq0}$, then the solution to $\dot{\hat{\rho}} = \ell_f \hat{\rho} + w_c$ is more conservative than the structure \eqref{eq:estimates}. 
\end{remark}

\begin{lemma} \label{lemma:observer_hybrid}
    Let $(t_1,j_1)$ be the instant after the first jump according to $\mathcal{D}_m$ in \eqref{eq:model_hybrid}-\eqref{eq:observer_hybrid}. Then the solutions to \eqref{eq:model_hybrid}-\eqref{eq:observer_hybrid} satisfy $\|r(t,j) - \hat{r}(t,j)\| \leq \hat{\rho}_r(t,j)$ and $\| v(t,j) - \hat{v}(t,j)\| \leq \hat{\rho}_v(t,j)$ for all $(t,j) \geq (t_1,j_1)$.
\end{lemma}

\extendedversion{
\begin{proof}
By Rule~\ref{rule:measurements}, the result holds at $(t_1,j_1)$ and immediately after every other jump according to $\mathcal{D}_a$. Thus, we only need to prove that the result is preserved along the flows and at jumps according to $\mathcal{D}_a$. Let $\tilde{r} = r - \hat{r}$ and $\tilde{v} = v - \hat{v}$. We now show that $\frac{d}{dt}\|\tilde{r}\| \leq \dot{\hat{\rho}}_r$ and $\frac{d}{dt}\|\tilde{v}\| \leq \dot{\hat{\rho}}_v$, from which it follows by the comparison lemma that the result holds during the flows.
\begin{flalign*}
    {\textstyle\frac{d}{dt}} \| \tilde{r} \| &= \| v - \hat{v} \| = \tilde{v} \leq \hat{\rho}_v = \dot{\hat{\rho}}_r \\
    {\textstyle\frac{d}{dt}} \| \tilde{v} \| &= \| F(t,x) - f(t,\hat{x}) \leq \| f(t,x) - f(t,\hat{x}) \| + w_c & \\
    &\leq \ell_{f,r} \|r - \hat{r}\| + \ell_{f,v} \| v - \hat{v} \| + w_c \\ &= \ell_{f,r} \|\tilde{r}\| + \ell_{f,v} \| \tilde{v} \| + w_c \leq \ell_{f,r} \hat{\rho}_r + \ell_{f,v} \hat{\rho}_v = \dot{\hat{\rho}}_v
\end{flalign*}
Next, Assumption~\ref{as:dynamics} implies $\|v^+ - v - g(t,u)\| \leq w_g(\|u\|)$. Thus,
\begin{flalign*}
    \| v^+ &- \hat{v}^+\| {=} \| v^+ - \hat{v} - g(t,u)\| {=} \| v^+ - v - g(t,u) + v - \hat{v}\| \\ &\leq \| v^+ - v - g(t,u)\| + \|v - \hat{v}\| \leq w_g(\|u\|) + \hat{\rho}_v = \hat{\rho}_v^+ 
\end{flalign*}
so the result also holds after each control application.
\end{proof}
}

Going forward, we also impose the following assumption.

\begin{assumption} \label{as:intersecting}
    Let $(t,j)$ and $(t,j+1)$ be the times immediately before and after a measurement instant. Assume that $\mathcal{B}_{\bar{\rho}_r(j+1)}(\bar{r}(j+1)) \subseteq \mathcal{B}_{\hat{\rho}_r(t,j)}(\hat{r}(t,j))$ and $\mathcal{B}_{\bar{\rho}_v(j+1)}(\bar{v}(j+1)) \subseteq \mathcal{B}_{\hat{\rho}_v(t,j)}(\hat{v}(t,j))$.
\end{assumption}

\noindent
By Lemma~\ref{lemma:observer_hybrid}, it must hold that $\mathcal{B}_{\hat{\rho}_r(t,j)}(\hat{r}(t,j)) \cap \mathcal{B}_{\bar{\rho}_r(j+1)}(\bar{r}(j+1))$ is nonempty. Assumption~\ref{as:intersecting} further requires that the entirety of the new set $\mathcal{B}_{\bar{\rho}_r(j+1)}(\bar{r}(j+1))$ lies within the previous uncertainty ball $\mathcal{B}_{\hat{\rho}_r(t,j)}(\hat{r}(t,j))$. This is not restrictive and can be enforced via measurement filtering.

\subsection{Prediction Functions} \label{sec:predictions}

Because the observer \eqref{eq:estimates} is entirely open-loop, the controller can predict what the state will be at any time in the future. To capture these predictions, define the function $p$ similarly to \cite[Eq.~(3)]{CDC23} as the solution, for $\tau \geq t$, to the following differential equation with initial condition $p(t,t,\hat{x}) = y(t) = x$,
\begin{equation}
    p(\tau,t,\hat{x})  
    \triangleq \begin{bmatrix} y_r(\tau) \\ y_v(\tau) \end{bmatrix} \textsr{ where } \begin{bmatrix} \dot{y}_r(s) \\ \dot{y}_v(s) \end{bmatrix} = \begin{bmatrix} y_v(s) \\ f(s,y(s)) \end{bmatrix} . \label{eq:path}
\end{equation}
That is, $p(\tau,t,\hat{x})$ is the future value of $\hat{x}$ if no jumps were to occur between $t$ and $\tau$. Similarly, following from \eqref{eq:linear_estimates}, denote
\begin{equation}
    q(\Delta,\hat{\rho}) \triangleq e^{A\Delta} \hat{\rho} + A^{-1}(e^{A\Delta} - I) [0,\,w_c]\transpose \label{eq:path_rho}
\end{equation}
with $A$ as in \eqref{eq:estimates}, to encode the solution for $\hat{\rho}$ without jumps. Recall that $\dot{\hat{\rho}}$ is time-invariant, so we drop \regularversion{argument $t$ in \eqref{eq:path_rho}}\extendedversion{the explicit dependence on argument $t$ in \eqref{eq:path_rho}}.

\begin{remark}
    If the system is subject to input delay, this can easily be captured through $p$, $q$, and $T_m$. Thus, we do not further address input delay here. \regularversion{See \cite[Remark~2]{extended_version}.}
    
    \extendedversion{That is, if there is a delay $T_d$, then introduce $\hat{x}^d = p(t+T_d,t,\hat{x})$ and $\hat{\rho}^d = q(T_d,\hat{\rho})$, and treat these delayed quantities as the state variables. Let $T_m^d = T_m + T_d$ replace $T_m$, and let $\bar{\rho}^d = q(T_d,\bar{\rho})$ replace $\bar{\rho}$ as the uncertainty at the soonest actionable instant after receiving a measurement.}
\end{remark}\vspace{-12pt}

\section{Control Law for Impulsive Actuators} \label{sec:safety_impulsive}

In this section, we develop conditions on the control input to guarantee forward invariance of a specified safe set under model \eqref{eq:model_hybrid}. Let $h:\mathcal{T}\times\mathcal{X}\rightarrow \reals$ be a given function. Suppose that the system is considered safe as long as $x(t) \in \mathcal{H}(t) \triangleq \{x\in\mathcal{X}(t) \mid h(t,x) \leq 0\}$. Since the state is uncertain, the value of $h$ is also uncertain, so we assume the following.

\begin{assumption} \label{as:h_lipschitz}
    Assume $h$ is locally Lipschitz continuous; i.e. there are known functions $\ell_{h,r}(t,x,\hat{\rho})$, $\ell_{h,v}(t,x,\hat{\rho})$ such that $|h(t,(r_1,v_1)) - h(t,(r_2,v_2))| \leq \ell_{h,r}(t,(r_1,v_1),\hat{\rho}) \| r_1 - r_2\| + \ell_{h,v}(t,(r_1,v_1),\hat{\rho}) \|v_1 - v_2\|, \forall r_2 \in\hspace{-1pt} \mathcal{B}_{\hat{\rho}_r}(r_1), v_2 \in\hspace{-1pt} \mathcal{B}_{\hat{\rho}_v}(v_1)$. Further, for any $x_1,x_2\in\mathcal{X}$ and associated $\hat{\rho}_1,\hat{\rho}_2$, if $\mathcal{B}_{\hat{\rho}_{1,r}}(r_1) \subseteq \mathcal{B}_{\hat{\rho}_{2,r}}(r_2)$ and $\mathcal{B}_{\hat{\rho}_{1,v}}(v_1) \subseteq \mathcal{B}_{\hat{\rho}_{2,v}}(v_2)$, assume 
    $\ell_{h,r}(t,x_1,\hat{\rho}_1) \leq \ell_{h,r}(t,x_2,\hat{\rho}_2)$ and $\ell_{h,v}(t,x_1,\hat{\rho}_1) \leq \ell_{h,v}(t,x_2,\hat{\rho}_2)$.
\end{assumption}

\noindent
That is, if one set $\mathcal{P}_1$ is entirely contained in a second set $\mathcal{P}_2$, then the Lipschitz constant on $\mathcal{P}_1$ is assumed less than the Lipschitz constant on $\mathcal{P}_2$. This is trivially true if the minimal Lipschitz constant is used. We thus have the bound
\begin{equation}
    h(t,x) \leq \underbrace{h(t,\hat{x}) + \ell_{h,r}(t,\hat{x},\hat{\rho})\hat{\rho}_r + \ell_{h,v}(t,\hat{x},\hat{\rho})\hat{\rho}_v}_{\triangleq \hat{h}(t,\hat{x},\hat{\rho})} . \label{eq:h_worst_case}
\end{equation}
which further satisfies Lemma~\ref{lemma:intersecting} below\regularversion{ (proven in \cite{extended_version})}. 
For brevity, define $\ell_h(t,\hat{x},\hat{\rho}) \triangleq [\ell_{h,r}(t,\hat{x},\hat{\rho}),\, \ell_{h,v}(t,\hat{x},\hat{\rho})]\transpose$.

\begin{lemma} \label{lemma:intersecting}
    Let $(t,j)$ and $(t,j+1)$ be the times immediately before and after a measurement instant. Let $\hat{h}$ be as in \eqref{eq:h_worst_case}. Then $\hat{h}(t,\hat{x}(t,j+1),\hat{\rho}(t,j+1)) \leq \hat{h}(t,\hat{x}(t,j),\hat{\rho}(t,j))$.
\end{lemma}

\extendedversion{
\begin{proof}
   For brevity, denote by $\hat{x}^-$, $\hat{\rho}^-$, and $\ell_h^-$ the quantities before the measurement, and by $\hat{x}^+$, $\hat{\rho}^+$, and $\ell_h^+$ the quantities after the measurement. 
    By Assumptions~\ref{as:intersecting}-\ref{as:h_lipschitz}, it immediately follows that $\ell_{h,r}^+ \leq \ell_{h,r}^-$, and similarly $\ell_{h,v}^+ \leq \ell_{h,v}^-$. 
    It also follows by Assumption~\ref{as:intersecting} that $\| \hat{r}^+ - \hat{r}^-\| \leq \hat{\rho}_r^- - \hat{\rho}_r^+$, and similarly $\| \hat{v}^+ - \hat{v}^-\| \leq \hat{\rho}_v^- - \hat{\rho}_v^+$. Thus,
    \begin{align*}
        \hspace{-0.25in}\hat{h}(t,\hat{x}^+,\hat{\rho}^+) &= h(t,\hat{x}^+) + \ell_{h,r}^+ \hat{\rho}_r^+ + \ell_{h,v}^+ \hat{\rho}_v^+ \\
        &\leq h(t,\hat{x}^+) + \ell_{h,r}^- \hat{\rho}_r^+ + \ell_{h,v}^- \hat{\rho}_v^+ \\
        &\leq [h(t,\hat{x}^-) + \ell_{h,r}^- (\hat{\rho}_r^- - \hat{\rho}_r^+) + \ell_{h,v}^- (\hat{\rho}_v^- - \hat{\rho}_v^+)] \nonumber \hspace{-0.25in} \\ &\;\;\;\;\;\; + \ell_{h,r}^- \hat{\rho}_r^+ + \ell_{h,v}^- \hat{\rho}_v^+ \\
        &= h(t,\hat{x}^-) + \ell_{h,r}^- \hat{\rho}_r^- + \ell_{h,v}^- \hat{\rho}_v^- \\ &= \hat{h}(t,\hat{x}^-,\hat{\rho}^-) \qedhere
    \end{align*}
\end{proof}
}

Next, proceeding similar to \cite{CDC23}, denote by $\psi_h$ any function satisfying\vspace{-10pt}
\begin{equation}
    \psi_h(\tau,t,\hat{x}) \geq h(s, p(s,t,\hat{x})) , \; \forall s\in[t,\tau] \,, \label{eq:psi_h} \hspace{-25pt}
\end{equation}
where \cite{CDC23} cites several possible methods to compute $\psi_h$. The method \cite[Eq.~(16)]{CDC23} is also applicable here via $\psi_h = \max [\psi_h^*]$.

To use \eqref{eq:h_worst_case} with \eqref{eq:psi_h}, we will also need a ``Lipschitz constant along a trajectory'', defined as
\begin{equation}
    \ell_{h,r}^p(\tau,t,\hat{x},\hat{\rho}) \triangleq  \max_{s\in[t,\tau]} \ell_{h,r}(s, p(s,t,\hat{x}), q(s-t,\hat{\rho})) \,. \label{eq:h_lipschitz_traj}
\end{equation}
Define $\ell_{h,v}^p$ similarly, and define $\ell_h^p$ similar to $\ell_h$ above.

We now have bounds on $h$ and $\ell_h$ in \eqref{eq:h_worst_case}, so next we bound $\hat{\rho}$. First, we define the set of \emph{guaranteed impulse opportunities}
\begin{equation}
    \boldsymbol{\Sigma}_a^\textsr{int} \triangleq \{ \sigma \in \boldsymbol{\Sigma} \mid \sigma_s = 0 \;\textrm{and}\; \sigma_a < 0 \;\textrm{and}\; \sigma_m > T_m\} \,.
\end{equation}
The set $\boldsymbol{\Sigma}_a^\textsr{int}$ is important because in that set, the controller's choice of $b$ entirely determines which jump in \eqref{eq:model_hybrid} occurs. That is, if $\sigma\in\boldsymbol{\Sigma}_a^\textrm{int}$, then $b = 1 \implies (b,\sigma) \in \mathcal{D}_a \setminus \mathcal{D}_c$ and $b=0 \implies (b,\sigma) \in \mathcal{D}_c \setminus \mathcal{D}_a$.

\extendedversion{
\begin{remark} \label{rem:lipschitz}
    Note that the assumption that $f$ is globally Lipschitz continuous on $\mathcal{X}$ in Assumption~\ref{as:dynamics} can be relaxed by instead redefining $\ell_{f,r}$ and $\ell_{f,v}$ as ``Lipschitz constants along a trajectory'' $\ell_{f,r}(\tau,t,\hat{x},\hat{\rho})$ and $\ell_{f,v}(\tau,t,\hat{x},\hat{\rho})$ similar to \eqref{eq:h_lipschitz_traj} with $\tau$ being the time of the next guaranteed impulse opportunity (see \eqref{eq:def_delta}). This would require that the matrix $A$ in \eqref{eq:path_rho} be recomputed each time that $q$ is invoked, and thus would increase computational complexity.
\end{remark}
}

Next, we assume that the minimum time between measurements $T_L$\extendedversion{{ }(see Rule~\ref{rule:measurements})} is sufficiently long that 
at least one guaranteed im\regularversion{-}pulse opportunity will occur during each measurement cycle:

\begin{assumption} \label{as:measurements}
    $T_L > T_m + T_s + \max\{T_a-T_m, 0\}$.
\end{assumption}

\extendedversion{
\begin{remark}
    The requirement in Assumption~\ref{as:measurements} can be relaxed if $T_L$ is small, though this would require that the times in \eqref{eq:q1}-\eqref{eq:def_delta} be recomputed. In such a case, the proper choice of these times will depend on if $T_M \leq T_a$ as well, so we do not provide specific formulas for these cases.
\end{remark}

}

Next, the longest possible delay between a measurement and an actuation is $T_M - T_m$. Assuming no prior actuations, the accumulated uncertainty over that time-span is \extendedversion{upper bounded by}\regularversion{less than}
\begin{equation}
    \hspace{-2pt} q_1 \smallop{=} e^{A (T_M-T_m)} [\rho_r, \hspace{0.25pt} \rho_v]\transpose + A^{-1} (e^{A (T_M-T_m)} - I)[0, \hspace{0.25pt} w_c]\transpose .\hspace{-3pt} \label{eq:q1}
\end{equation}
Next, the maximum number of actuations during that time-span is $N = 1 + \floor((T_M - T_m)/T_a)$. Excluding the last actuation impulse, the maximal uncertainty due to the impulses over this time-span is \extendedversion{upper bounded by}\regularversion{less than}
\begin{equation}
    \textstyle q_2 = \sum_{i=1}^{N-1}e^{A(T_M - T_m - T_a (i-1))} [0,\, W_g]\transpose \,. \label{eq:q2}
\end{equation}
By linearity of \eqref{eq:estimates}, we can simply add these two uncertainties to define $q_3 \triangleq q_1 + q_2$, the maximal uncertainty that will ever be present \emph{before} an actuation.

We can thus state a definition of CBF as follows.

\begin{definition} \label{def:impulsive_cbf}
    Let $\psi_h$ satisfy \eqref{eq:psi_h} and let $q_3=q_1+q_2$ be as in \eqref{eq:q1}-\eqref{eq:q2}. Let $\delta_r = T_a + T_m + 2T_s$. A continuous function $h$ is a \emph{Robust Impulsive Timed CBF (RIT-CBF)} if 
    \begin{multline}
        \hspace{-6pt} 0 \geq \inf_{u\in\mathcal{U}} [\psi_h(t + \delta_r, t, \hat{x}^+) + \ell_h^p(t+\delta_r, t, \hat{x}^+, \hat{\rho}^+)\transpose q(\delta_r,\hat{\rho}^+)] \\
        \textsr{with } \hat{x}^+ = [\hat{r}, \hat{v} + g(t,u)]\transpose,\, \hat{\rho}^+ = \hat{\rho} + [0,\,w_g(\|u\|)]\transpose \hspace{-4pt} \label{eq:cbf_definition_impulsive}
    \end{multline}
    for all $(\hat{x}, \hat{\rho}) \in \{ x\in\mathcal{X}(t), \hat{\rho} \leq q_3 \mid \hat{h}(t,\hat{x},\hat{\rho}) \leq 0\}, t\in\mathcal{T}$.
\end{definition}

That is, $h$ is an RIT-CBF if at any state $(\hat{x},\hat{\rho})$ and time $t${} such that $\hat{h}(t,\hat{x},\hat{\rho})\leq 0$, one can find a control input such that $\hat{h}$ at the next impulse opportunity, which will always occur before $t+\delta_r$, can be made nonpositive. Intuitively, this means that the reduction in $h$ must outpace the increase in uncertainty $\ell_h^p(\cdot)\transpose q(\cdot)$. Since uncertainty grows exponentially, this implies a limit on the maximal allowable $T_M$. 

If $h$ is an RIT-CBF, then we know that $\mathcal{H}$ is a ``good'' choice of safe set, i.e. a safe set that can be rendered forward invariant under \eqref{eq:model_hybrid}. Condition{} \eqref{eq:cbf_definition_impulsive} can be verified offline, and if it is satisfied, then Theorem~\ref{thm:impulsive_safety} below provides an online condition that the control law should satisfy to guarantee that trajectories remain within $\mathcal{H}$. Similar to \cite[Thm.~1]{CDC23}, Theorem~\ref{thm:impulsive_safety} provides two possible conditions. 
Suppose that at time $(t,j)$ the timers $\sigma$ are such that the controller block in Fig.~\ref{fig:block_diagram} is activated. If the ``system checks'' in Fig.~\ref{fig:block_diagram} selects $b=0$ and applies no impulse, then the next guaranteed impulse opportunity will occur either exactly at $t+T_s$, or if $\sigma_m$ is close to $T_m$, then sometime during $[t+\sigma_m, t + \sigma_m + T_s]$. Thus, define the horizon%
\begin{subequations}\begin{flalign}
    \delta_1(\sigma_m) &= \begin{cases}
        \sigma_m + T_s & \textsr{if }\sigma_m \leq T_m + T_s \\
        T_s & \textsr{otherwise.}
\end{cases} &\label{eq:def_delta1}
\end{flalign}
That is, the system can save fuel by coasting with no actuation if the ``system checks'' predicts that the system will remain safe at least until $t+\delta_1$, which is verified by \eqref{eq:cbf_condition_impulsive_passive} below.

Alternatively, if the controller selects $b=1$, then the next guaranteed impulse opportunity will occur sometime during $(t+T_a, t+T_a+T_s]$, or if $\sigma_m$ is close to $T_m$, then by Assumption~\ref{as:measurements}, such an opportunity will occur within $T_s$ of the latter of $t+T_a$ and $t+\sigma_m$. Thus, define the horizon
\begin{flalign}
    \delta_2(\sigma_m) &\smallop{=} \begin{cases}
        \max\{\sigma_m,T_a\} + T_s & \textsr{if }\sigma_m \smallop{\leq} T_a \smallop{+} T_m \smallop{+} T_s \\
        T_a + T_s & \textsr{otherwise.}
        \end{cases} \hspace{-18pt} &  \label{eq:def_delta2}
\end{flalign}\label{eq:def_delta}\end{subequations}
In this case, the ``nonlinear program'' block of Fig.~\ref{fig:block_diagram} is activated to compute a control input satisfying \eqref{eq:cbf_condition_impulsive_active} below.

\begin{theorem} \label{thm:impulsive_safety}
    Let $(t_0,0)$ be the initial time. Suppose that $\sigma_m(t_0,0) = 0$, $\sigma_m(t_0,1) > 0$, $\sigma_a(t_0,1) <0$, $\sigma_s(t_0,1) = 0$, and $\hat{h}(t_0,\bar{x}(0),\bar{\rho}(0)) \leq 0$. That is, $(t_0,0)$ is a measurement instant and $(t_0,1)$ is a guaranteed impulse opportunity.
    If the control inputs $b,u$ are such that either A) $b=0$ and
    \begin{subequations}
    \begin{equation}
        0 \geq \psi_h(t+\delta_1, t, \hat{x}) + \ell_h^p(t+\delta_1, t, \hat{x}, \hat{\rho})\transpose q(\delta_1,\hat{\rho}) \label{eq:cbf_condition_impulsive_passive}
    \end{equation}
    or B) $b=1$ and
    \begin{multline}
        \hspace{-8pt} 0 \geq \psi_h(t+\delta_2, t, \hat{x}^+) + \ell_h^p(t+\delta_2,t,\hat{x}^+,\hat{\rho}^+)\transpose  q(\delta_2,\hat{\rho}^+)
        \\
        \textsr{with } \hat{x}^+ = [\hat{r}, \hat{v} + g(t,u)]\transpose,\, \hat{\rho}^+ = \hat{\rho} + [0,\,w_g(\|u\|)]\transpose \hspace{-4pt} \label{eq:cbf_condition_impulsive_active}
    \end{multline}\label{eq:cbf_condition_impulsive}%
    \end{subequations}
    for all $\sigma\in\boldsymbol{\Sigma}_a,\hat{x}\in\mathcal{X}(t),\hat{\rho}\leq q_3,t\in\mathcal{T}$, then $x(t,j) \smallin \mathcal{H}(t)$ for all $(t,j)\geq (t_0,1)$.
\end{theorem}

\regularversion{\noindent{}The proof follows by construction of \eqref{eq:def_delta}-\eqref{eq:cbf_condition_impulsive}, as well as Lemma~\ref{lemma:intersecting}, and is included in \cite[Thm.~1]{extended_version}.}

\extendedversion{\begin{proof}
    The proof follows by construction. As explained above, if $(t_j,j)$ is an impulse opportunity, then $\delta_1$ and $\delta_2$ are upper bounds on $t_k - t_j$ until the next time $(t_k,k)$ that the controller is guaranteed another impulse opportunity. Note that $\psi_h$, $\ell_h^p$ and $q$ are all nondecreasing in their first argument, so it follows by the constructions \eqref{eq:h_worst_case}-\eqref{eq:psi_h} that \eqref{eq:cbf_condition_impulsive_passive} and \eqref{eq:cbf_condition_impulsive_active} are upper bounds on $\hat{h}$ over the interval either a) $[t,t+\delta_1]$ if $b=0$ or b) $[t,t+\delta_2]$ if $b=1$. Thus, \eqref{eq:cbf_condition_impulsive_passive}-\eqref{eq:cbf_condition_impulsive_active} guarantee that $\hat{h}(t,\hat{x}(t,j),\hat{\rho}(t,j))\leq 0$ over those intervals, which each must contain another guaranteed impulse opportunity. 
    Additionally, if those intervals contain a measurement instant, then Lemma~\ref{lemma:intersecting} guarantees that $\hat{h}$ remains nonpositive after the measurement instant. 
    Since $\sigma_m(t_0,1) \in \boldsymbol{\Sigma}_a^\textrm{int}$ and \eqref{eq:cbf_condition_impulsive_passive}-\eqref{eq:cbf_condition_impulsive_active} hold at every future control computation, it follows by induction that $\hat{h}(t,\hat{x}(t,j),\hat{p}(t,j))\leq 0$ for all $(t,j) \geq (t_0,1)$. Since $\hat{h}$ is an upper bound for $h$, the result follows.
\end{proof}}

That is, any control law that satisfies \eqref{eq:cbf_condition_impulsive} will guarantee that trajectories always remain in the safe set $\mathcal{H}$. Frequently, \eqref{eq:cbf_condition_impulsive} is enforced using optimization-based control laws such as \cite[Sec.~IIc]{CBFs_Tutorial} or \cite[Eq.~(18)]{CDC23}. Note that the quantity $\delta_r$ in Definition~\ref{def:impulsive_cbf} is an upper bound on $\delta_2$ in \eqref{eq:def_delta}, so if $h$ is an RIT-CBF, then such optimizations will be recursively feasible. Thus, we have extended our prior work \cite{CDC23} to perturbed models and measurements.

\extendedversion{
\begin{remark}
    In our prior work \cite{CDC23}, we also studied asymptotic stability conditions. We exclude such a study here, but conjecture that the usual result that exponential stability (i.e. $V(t + T) \leq \alpha(V(t))$ with linear $\alpha$ function) implies input-to-state stability will hold for the dynamics \eqref{eq:model_hybrid} as well. 
\end{remark}
}

\section{Control Law for Continuous Actuators} \label{sec:continuous}

In this section, we suppose the satellite possesses only continuously-applied actuators (or sampled actuators with sampling period $T_s \ll T_M$ that are approximated as continuous). With these actuators, the model \eqref{eq:model_hybrid} simplifies to
\begin{equation} \hspace{-6pt}
    \begin{cases}
        \dot{r} = v \\ 
        \dot{v} \in C(t,x,u) & \sigma_m \geq 0 \\
        \dot{\sigma}_m = -1
    \end{cases} , \hspace{8pt} \begin{cases}
        r^+ = r \\
        v^+ = v & \sigma_m = 0, \\
        \sigma_m^+ = \bar{\sigma}_m(j)
    \end{cases} \label{eq:model_continuous} \hspace{-6pt}
\end{equation}
where $C(t,x,u) \triangleq F(t,x) + G(t,u)$ and Assumptions~\ref{as:dynamics}-\ref{as:intersecting}{} and Rule~\ref{rule:measurements} still apply. Assume that $F$, $G$, and $u$ are sufficiently regular that solutions to \eqref{eq:model_continuous} exist for all times $t\in\mathcal{T}$. We construct an open-loop observer similar to \eqref{eq:observer_hybrid} as
\begin{equation} \hspace{-4pt}
    \begin{cases}
        \dot{\hat{r}} = \dot{\hat{v}} \\
        \dot{\hat{v}} = f(t,\hat{x}) + g(t,u) \\
        \dot{\hat{\rho}}_r = \hat{\rho}_v \\
        \dot{\hat{\rho}}_v = \ell_{f,r}\hat{\rho}_r + \ell_{f,v} \hat{\rho}_v + w_c + w_g(\|u\|)
    \end{cases} \hspace{-6pt} \begin{cases}
        \hat{r}^+ = \bar{r}(j) \\
        \hat{v}^+ = \bar{v}(j) \\
        \hat{\rho}_r^+ = \bar{\rho}_r(j) \\
        \hat{\rho}_v^+ = \bar{\rho}_v(j)
    \end{cases} \hspace{-12pt} \label{eq:observer_continuous}
\end{equation}
where the flow and jump sets are the same as in \eqref{eq:model_continuous}.

\begin{lemma} \label{lemma:observer_continuous}
    Let $(t_1,j_1)$ be the instant after the first jump in \eqref{eq:model_continuous}-\eqref{eq:observer_continuous}. Then the solutions to \eqref{eq:model_continuous}-\eqref{eq:observer_continuous} satisfy $\|r(t,j) - \hat{r}(t,j)\| \leq \hat{\rho}_r(t,j)$ and $\| v(t,j) - \hat{v}(t,j)\| \leq \hat{\rho}_v(t,j)$ for all $(t,j)\geq (t_1,j_1)$.
\end{lemma}

\extendedversion{
\begin{proof}
    The proof is identical to the proof of Lemma~\ref{lemma:observer_hybrid} with a slightly different formula for $\frac{d}{dt}\|\tilde{v}\|$ as follows
    \begin{align*}
        {\textstyle\frac{d}{dt}}\|\tilde{v}\| &= \| F(t,x) + G(t,u) - f(t,\hat{x}) - g(t,u)\| \\ &\leq \| F(t,x) - f(t,\hat{x})\| + \| G(t,u) - g(t,u)\| \\ &\leq \| f(t,x) - f(t,\hat{x})\| + w_c + w_g(\|u\|) \\ &\leq \ell_{f,r} \| r - \hat{x} \| + \ell_{f,v} \| v - \hat{v} \| + w_c + w_g(\|u\|) \\ &\leq \ell_{f,r} \hat{\rho}_r + \ell_{f,v} \hat{\rho}_v + w_c + w_g(\|u\|) = \dot{\hat{\rho}}_v 
    \end{align*}
    Note that one could also generalize \eqref{eq:model_continuous}-\eqref{eq:observer_continuous} to $G$ that are functions of $x$ as well, just with a longer expression for $\dot{\hat{\rho}}_v$.
\end{proof}
}

Since the control input acts at every instant, there is no need for the prediction functions $p$ and $\psi_h$ in the prior section. We now assume that $T_m = 0$ and $T_L > 0$.
Since the observer \eqref{eq:observer_continuous} is slightly different from \eqref{eq:observer_hybrid}, the formula for the predicted error is now given by $q^*$ as below
\begin{equation}
    \hspace{-4pt} \hat{\rho}(t+\Delta,j) \leq \underbrace{e^{A \Delta} \hat{\rho}(t,j) + A^{-1} (e^{A \Delta} - I) [0, w_c + W_g]\transpose}_{\triangleq q^*(\Delta, \hat{\rho}(t,j))} \hspace{-4pt} \label{eq:continuous_time_q}
\end{equation}
and the maximal uncertainty that could occur between a measurement and actuation is now $q_4 \triangleq q^*(T_M, (\rho_r, \rho_v))$.

Let $h(t,x)$ and $\mathcal{H}(t)$ be as in Section~\ref{sec:safety_impulsive}, and suppose the following assumption, from which Definition~\ref{def:continuous_cbf} follows.

\begin{assumption}
    Let Assumption~\ref{as:h_lipschitz} hold. Further assume that $\ell_h(\cdot) = [\ell_{h,r}(\cdot),\,\ell_{h,v}(\cdot)]\transpose$ are continuously differentiable.
\end{assumption}

\begin{definition} \label{def:continuous_cbf}
    Let $\hat{h}$ be as in \eqref{eq:h_worst_case} and $q_4= q^*(T_M, (\rho_r, \rho_v))$ as in \eqref{eq:continuous_time_q}. A continuously differentiable function $h$ is a \emph{Robust Timed CBF (RT-CBF)} if there exists $\alpha\in\mathcal{K}$ such that
    \begin{equation}
        \inf_{u\in\mathcal{U}} {\textstyle\frac{d}{dt}}[\hat{h}](t,\hat{x},\hat{\rho},u) \leq \alpha(-\hat{h}(t,\hat{x},\hat{\rho})) \label{eq:cbf_definition_continuous}
    \end{equation}
    for all $(\hat{x},\hat{\rho})\in\{x\in\mathcal{X}(t),\hat{\rho}\leq q_4 \mid \hat{h}(t,\hat{x},\hat{\rho}) \leq 0\}, t \in \mathcal{T}$.
\end{definition}

Similar to Definition~\ref{def:impulsive_cbf}, $h$ is an RT-CBF if the reduction in $h$ can always outpace the increase in uncertainty. This will only hold up to some maximal time $T_M$. Similar to the prior section, condition \eqref{eq:cbf_definition_continuous} can be checked offline. Then the online control law should satisfy the following condition.

\begin{theorem}
    Let $(t_0,0)$ be the initial time. Suppose that $\sigma_m(t_0,0) \smallop{=} 0$, $\sigma_m(t_0,1) \smallop{>} 0$, and $\hat{h}(t_0,\bar{x}(0),\bar{\rho}(0)) \smallop{\leq} 0$. Let $\mathcal{O}$ be an open set satisfying $\mathcal{O} \supset \mathcal{T}\times\mathcal{H}\times\{\hat{\rho}\in\reals^2\mid 0 \leq \hat{\rho} \leq q_4\}$.

    \noindent
    If there exists $\alpha\in\mathcal{K}$ such that the control law $u$ satisfies
    \begin{equation}
        {\textstyle\frac{d}{dt}}[\hat{h}](t,\hat{x},\hat{\rho},u) \leq \alpha(-\hat{h}(t,\hat{x},\hat{\rho})) \label{eq:cbf_condition_continuous}
    \end{equation}
    for all $(t,\hat{x},\hat{\rho})\in\mathcal{O}$, then $x(t,j) \in \mathcal{H}(t)$ for all $(t,j)\geq (t_0,1)$.
\end{theorem}

\regularversion{\noindent{}The proof follows from typical CBF theory involving \eqref{eq:cbf_condition_continuous}, along with Lemma~\ref{lemma:intersecting}, and is included in \cite[Thm.~2]{extended_version}.}

\extendedversion{
\begin{proof}
By Lemma~\ref{lemma:observer_continuous}, $\hat{h}$ is an upper bound on $h$, so it follows that $x(t_0,1) \in \mathcal{H}(t_0)$. Moreover, by \cite[Lemma~1]{CDC22} the conditions $\phi(t_0) \leq 0$ and $\dot{\phi} \leq \alpha(-\phi)$ on any open set $\boldsymbol{\Phi} \subset\reals$ satisfying $0\in\boldsymbol{\Phi}$ are sufficient to ensure $\phi(t) \leq 0$ for all $t\geq t_0$. Thus, by the comparison lemma \cite[Lemma IX.2.6]{nonlipschitz}, \eqref{eq:cbf_condition_continuous} is sufficient to render $\hat{h}(t,\hat{x}(t,j),\hat{\rho}(t,j))\leq 0$ for all $(t,j)$ until the next measurement. By Lemma~\ref{lemma:intersecting}, $\hat{h}$ will remain nonpositive after the measurement. Condition \eqref{eq:cbf_condition_continuous} then ensures that $\hat{h}$ remains nonpositive until the next measurement, so by induction, the result holds for all $(t,j) \geq (t_0,1)$.
\end{proof}
}

Similar to Section~\ref{sec:safety_impulsive}, we often enforce \eqref{eq:cbf_condition_continuous} using optimization-based control laws, and if $h$ is an RT-CBF, then such control laws will be recursively feasible. Thus, we have extended Sections~\ref{sec:impulsive_model}-\ref{sec:safety_impulsive} to continuously-applied actuators too.

\section{Simulation Case Studies} \label{sec:simulations}

\begin{figure}
    \centering
    \includegraphics[width=0.49\columnwidth,trim={0.7cm, 0cm, 1.8cm, 0.75cm},clip]{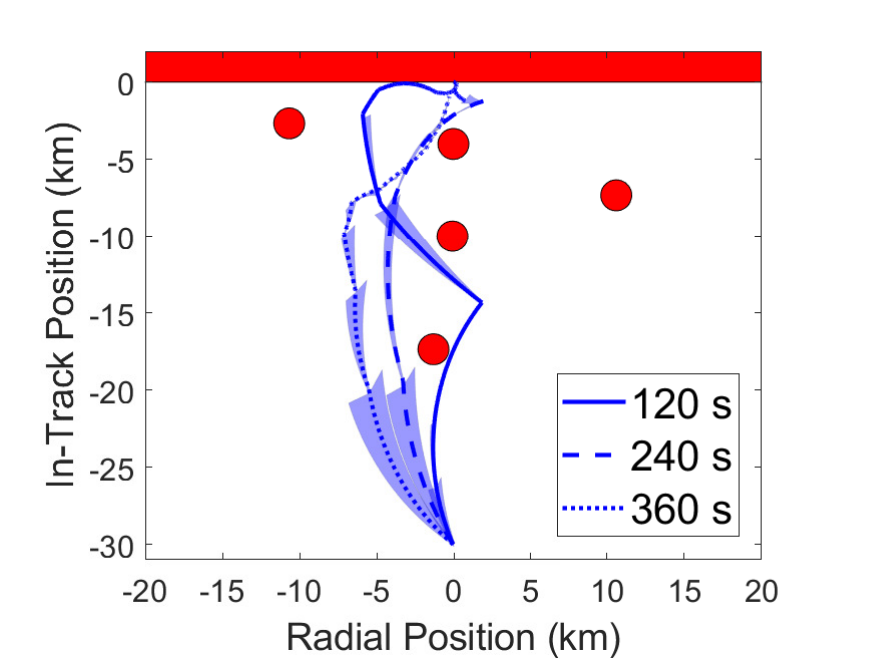}
    \hfill\resizebox{0.484\columnwidth}{!}{\begin{tikzpicture}
        \node[anchor=south west,inner sep=0] (image) at (0,0) {\includegraphics[width=0.85\columnwidth,trim={0cm, 0.8cm, 4.5cm, 1.5cm},clip]{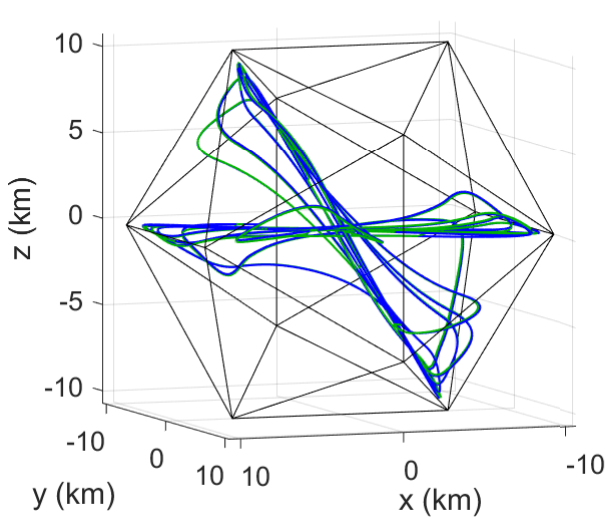}};
        \node[anchor=south west,inner sep=0] (image) at (4.8,5.3) {\includegraphics[width=0.3\columnwidth,trim={10.77cm, 9.06cm, 1.05cm, 0.83cm},clip]{pos_plot_continuous2.eps}};
    \end{tikzpicture}}
    \caption{Left: Plots of trajectories and uncertainties for the impulsive case study for various $T_M$, plotted in a rotating frame centered at the target, with the $x$ axis always along the target's radial direction. Note that the obstacles perform ellipses about the origin and are not static in this frame. Right: Plots of the real and estimated trajectories of the continuous case study inside a specified icosahedron.}
    \label{fig:positions}
\end{figure}

\begin{figure}
    \centering
    \includegraphics[width=0.94\columnwidth,trim={0.6cm, 0cm, 1.1cm, 0.1cm},clip]{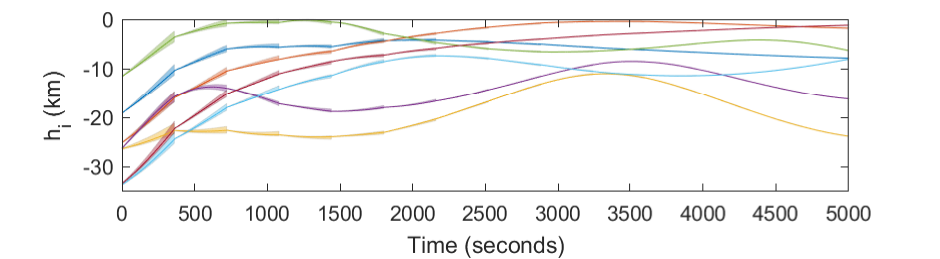}\\
    \vspace{2pt}
    \includegraphics[width=0.94\columnwidth,trim={0.6cm, 0cm, 1.4cm, 0.3cm},clip]{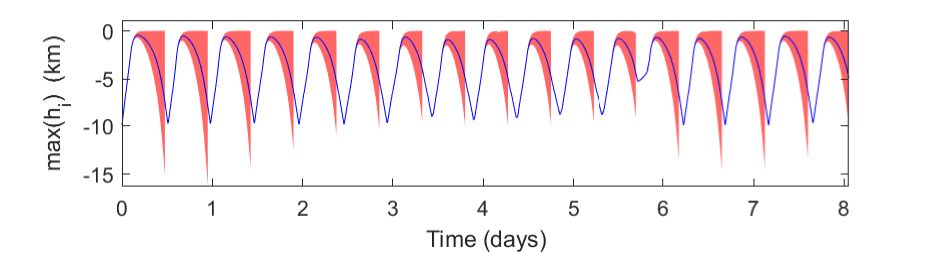}
    \caption{Top: Plot of the 7 CBF values and estimates (shaded) for $T_M = 360$~s in the impulsive case study. Bottom: Plot of the max of the 20 CBF values and estimates in the continuous case study.}
    \label{fig:h_values}
\end{figure}

We verify the strategies of Sections~\ref{sec:safety_impulsive}-\ref{sec:continuous} in simulation. For the impulsive model \eqref{eq:model_hybrid}, we use the example studied in \cite[Sec.~IV]{CDC23}. Let $r,v,u\in\reals^2$ and let $\mu\in\reals_{>0}$ be a constant. Let
\begin{equation}
     f(t,x) = -\mu r / \|r\|^3,  \; g(t,u) = u \,. 
    \label{eq:fg}
\end{equation}
Specifically, we consider a chaser satellite performing rendezvous with a target satellite in a elliptical orbit with semi-major axis of 7775~km and eccentricity of 0.1. Let $\mathcal{X}$ be a neighborhood of this nominal orbit;{} in this regime, $\ell_{f,r} = 0.000921$ and $\ell_{f,v} = 0$. Let $w_c = 9.2(10)^{-6}$~m/s$^2$ and $w_g(\lambda) = \min\{0.05\lambda, 5\textsr{ m/s}\}$. For the CBF, we use the same $h,\psi_h$ as in \cite[Eqs.~(14),(16)]{CDC23} with $\gamma = 1$ and $\delta = T_M / 12$ for 7 time-varying exclusion zones (7 CBFs) and we use the same optimization-based control law as in \cite[Eq.~(18b)]{CDC23}, now with $\hat{h}$ as in \eqref{eq:h_worst_case} in place of $\psi_h$ in \cite[Eq.~(18e)]{CDC23}. We also did not change any of the control Lyapunov function work from \cite{CDC23}. For simplicity, we programmed this control law to always perform one actuation per measurement cycle and chose $T_L = T_M$. More details can be found in the simulation code below\footnote{\texttt{\url{https://github.com/jbreeden-um/phd-code/tree/main/2024/LCSS\%20Robust\%20Hybrid}}}. We then simulated the above control law for various choices of $T_M$. Plots of the real trajectories and uncertainty regions (shaded zones) for three choices of $T_M$ are shown in Fig.~\ref{fig:positions} and a video showing the complete orbital dynamics is also available at \texttt{\url{https://youtu.be/ZVhYwji5Qg8}}. Finally, the CBF values $h(t,x)$ and estimated bounds $h(t,\hat{x}) \pm \ell_h(\cdot)\transpose q(\cdot)$ for the longest of these three simulations is shown in Fig.~\ref{fig:h_values}.

In summary, we were able to generalize our prior work in \cite{CDC23} to perturbed dynamics with the measurement model \eqref{eq:observer_hybrid}. For these disturbances, the longest measurement interval for which this choice of $h$ is an RIT-CBF is 390~s, whereas the controller was able to handle much longer periods between actuations in the case without disturbances in \cite{CDC23}.

Next, we tested the continuous actuation model \eqref{eq:model_continuous}. The $f$ and $g$ functions are still as in \eqref{eq:fg}, now with $w_c = 4.56(10)^{-6}$~m/s$^2$, which is the solar radiation pressure on a 10~kg/m$^2$ object at Earth, and with $w_g(\lambda) = \min\{0.02\lambda, 0.00005\textsr{ m/s}^2\}$,~$\bar{\rho}_r = 5$~m, and $\bar{\rho}_v = 5$~mm/s. Though $f$ and $g$ are the same, instead of satellite rendezvous in Low Earth Orbit, we now consider stationkeeping in a geosynchronous orbit. Now, the target satellite is a ``virtual target'', and the chaser satellite is commanded to stay within a 10~km sphere of this target. This sphere is approximated by an inscribed 20-sided polytope, yielding 20 constraints $\kappa_i(r) = p_i\transpose r - \rho$ and 20 CBFs $h_i(x) = \kappa_i(r) + \gamma \dot{\kappa}_i(r,v)$ with $\gamma = 120$. We choose the~control
\begin{equation}
    u = \argmin_{u\in\reals^3} \|u\|^2 \textrm{ s.t. } {\textstyle\frac{d}{dt}}[\hat{h}_i](t,x,u) \leq \alpha(-\hat{h}_i(t,x))\, \forall i
\end{equation}
for $\alpha(\lambda) = 0.004 \lambda$. Selecting $T_M = 11.4$~hours, the real and estimated trajectories are shown inside the specified polytope in Fig.~\ref{fig:positions} (see also \texttt{\url{https://youtu.be/7MeRHWCWc8E}}) and the maximal of the 20 CBF values and the corresponding estimates are shown in Fig.~\ref{fig:h_values}. We have tuned $\gamma$ so that the maximal $u$ applied is $8(10)^{-4}$~m/s$^2$, thus maintaining the low-thrust regime. Note that for these disturbances and this choice of $h_i$, 8.18~hours is the upper limit on allowable $T_M$ for which these $h_i$ functions are RT-CBFs, but the simulation still worked up to 11.4~hours because of the conservatism in $W_g$ in \eqref{eq:continuous_time_q}.

\section{Conclusions} \label{sec:conclusions}

This paper has presented set invariance conditions for second-order systems with infrequent measurements and open-loop observers. We started by generalizing our prior work with impulsive actuators to non-deterministic dynamics, and then developed similar results for continuously-applied low-thrust actuators. These advances make CBF theory more relevant to perturbed dynamics such as those in our simulations. In the future, we are interested in possible extensions of these results to satellite attitude dynamics, and to analyzing the stability properties of systems subject to this hybrid dynamical model.

\regularversion{\vspace{-3pt}}

\bibliographystyle{ieeetran}
\bibliography{sources}

\end{document}